\documentclass[pra,preprint,showpacs,showkeys,amsfonts]{revtex4}
\RequirePackage{times}
\RequirePackage{mathptm}
\begin{document}

\title{Eutactic quantum codes}
\author{Karl Svozil}
 \email{svozil@tuwien.ac.at}
\homepage{http://tph.tuwien.ac.at/~svozil}
\affiliation{Institut f\"ur Theoretische Physik, University of Technology Vienna,
Wiedner Hauptstra\ss e 8-10/136, A-1040 Vienna, Austria}

\begin{abstract}
We consider sets of quantum observables corresponding to {\em eutactic stars}. Eutactic stars are systems of vectors which are the lower-dimensional ``shadow'' image, the orthogonal view, of higher-dimensional orthonormal bases. Although these vector systems are not comeasurable, they represent redundant coordinate bases with remarkable properties. One application is quantum secret sharing.
\end{abstract}


\maketitle

The increased experimental feasibility to manipulate single or few particle quantum states, and
the theoretical concentration on the algebraic properties of the mathematical models
underlying quantum mechanics have stimulated a wealth
of applications in information and computation theory
\cite{Gruska,nielsen-book}.
In this line of reasoning, we shall consider quantized systems
which are in a coherent superposition of constituent states
in such a way that only the coherent superposition of these pure states
is in a predefined state; whereas one or all of the constituent states
are not.
Heuristically speaking, only the coherently combined states yield
the ``encoded message,'' the constituents or ``shares'' do not.

This feature could be compared to
``quantum secret sharing'' schemes
\cite{hil-bu-be-1999,gotesman-1999,gotesman-2000,ter-Divinc-leu-2001,Div-hay-ter-2002},
as well as to ``entangled entanglement´´
scenarios \cite{krenn1,cereceda-pra97}.
There, mostly entangled  multipartite system are investigated.
Thus, while the above cases concentrate mainly on quantum entanglement,
in what follows quantum coherence will be utilized:
in the secret-sharing scheme proposed here,
one party receives part of a quantum state and the other party receives the other part.
The parts are components of a vector lying in subspaces of a higher-dimensional
Hilbert space.
While the possible quantum states to be sent are orthogonal,
the parts are not,
so that the parties must put their parts together to decipher the message.

We shall deal with the general case first and consider examples later.
Consider an orthonormal basis ${\cal E} = \left\{ {\bf e}_1,\ldots ,{\bf e}_n \right\}$ of
the $n$-dimensional real Hilbert space ${\mathbb R}^n$ [whose origin is at $(0,\ldots ,0)$].
Every point ${\bf x}$  in ${\mathbb R}^n$
has a coordinate representation $x_i=\langle {\bf x}\mid {\bf e}_i\rangle$, $i=1,\ldots , n$
with respect to the basis ${\cal E}$.
Hence, any vector from the origin ${\bf v}={\bf x}$ has a representation in terms of the
basis vectors given by
${\bf v}=\sum_{i=1}^n \langle {\bf v}\mid {\bf e_i}\rangle {\bf e_i}={\bf v} \sum_{i=1}^n  [{\bf e_i}^T{\bf e_i}]$,
where the matrix notation has been used, in which ${\bf e_i}$ and ${\bf v}$
are row vectors and "$^T$" indicates transposition.
($\langle \cdot \mid \cdot \rangle$ and the matrix $[{\bf e_i}^T{\bf e_i}]$
stands for the scalar product and the dyadic product
of the vector ${\bf e_i}$ with itself, respectively).
Hence, $\sum_{i=1}^n  [{\bf e_i}^T{\bf e_i}]= {\mathbb I}_n$,
where ${\mathbb I}_n$ is the $n$-dimensional identity matrix.

Next, consider more general, redundant, bases consisting of systems of ``well-arranged''
linear dependent vectors
${\cal F} = \left\{ {\bf f}_1,\ldots ,{\bf f}_m \right\}$ with $m>n$,
which are the orthogonal projections of orthonormal bases of $m$-
(i.e., higher-than-$n$-) dimensional Hilbert spaces.
Such systems are are often referred to as {\em eutactic stars}
\cite{schlaefli-1901,hadwiger-40,coxeter-polytopes,seidel-76,hasse-stachel96}.
When properly normed, the sum of the dyadic products of their vectors yields unity;
i.e., $\sum_{i=1}^m  [{\bf f_i}^T{\bf f_i}]= {\mathbb I}_n$, giving raise to redundant
eutactic coordinates $x'_i=\langle {\bf x}\mid {\bf f}_i\rangle $, $i=1,\ldots , m>n$.
Indeed, many properties of operators and tensors defined
with respect to standard orthonormal bases directly translate into eutactic coordinates
\cite{hasse-stachel96}.

In terms of $m$-ary (radix $m$) measures of quantum information based on  state partitions
\cite{svozil-2002-statepart-prl},
$k$ elementary $m$-state systems can carry $k$ {\em nits}
\cite{zeil-99,zeil-bruk-99,DonSvo01}.
A nit can be encoded by the one-dimensional subspaces of ${\mathbb R}^m$
spanned by some orthonormal basis vectors ${\cal E}'=\left\{ {\bf e}_1,\ldots ,{\bf e}_m \right\}$.
In the  quantum logic approach pioneered by Birkhoff and von Neumann
(e.g., \cite{birkhoff-36,v-neumann-49,mackey:63,svozil-ql}),
every such basis vector corresponds to the physical proposition that ``the system is in a
particular one of $m$ different states.'' All the propositions corresponding to orthogonal
base vectors are comeasurable.

On the contrary, the propositions corresponding to the
eutactic star
$${\cal F}=\left\{ P{\bf e}_1,\ldots ,P{\bf e}_m \right\}$$
formed by some orthogonal projection $P$
of ${\cal E}'$
is no longer comeasurable (or it just spans a one-dimensional subspace).
Neither is the
eutactic star
$${\cal F}^\perp =\left\{ P^\perp {\bf e}_1,\ldots ,P^\perp {\bf e}_m \right\}$$
formed by the orthogonal projection $P^\perp $
of ${\cal E}'$.
Indeed, the elements of
${\cal F}$  and
${\cal F}^\perp$
may be considered as ``shares'' in the context of quantum secret sharing.
Thereby, not all shares may be equally suitable for cryptographic purposes.
This scenario can be generalized to multiple shares in a straightforward way.

Let us consider an example for a two-component two-share configuration, in
which each party obtains one substate from two possible ones.
In particular, consider
the two shares
$\{ {\bf w},{\bf x}\}$
and
$\{ {\bf y},{\bf z} \}$
defined in fourdimensional complex Hilbert space by
\begin{equation}
\begin{array}{cccc}
  {\bf w}= \left( 0,0,-\frac{1}{2\,{\sqrt{2}}}  ,{1\over \sqrt{2}}\right),\quad &
  {\bf x}=\frac{1}{2} \left( 0,0, -\frac{3}{2} ,-1\right),  \\
  {\bf y}=\frac{1}{2} \left(\frac{1}{{\sqrt{2}}},-1,0,0\right),\quad &
  {\bf z}=\frac{1}{2{\sqrt{2}}} \left(  -\frac{1}{{\sqrt{2}}}  ,  -1 ,0,0\right),  \\
\end{array}
\label{2003-eu-el1}
\end{equation}
While $\{ {\bf w} , {\bf x}\}$ and $\{{\bf y},{\bf z} \}$
constitute eutactic stars in ${\mathbb R}^2$,
the coherent superposition of
${\bf w}$ with  ${\bf y}$,
and ${\bf x}$   with  ${\bf z}$
yield two orthogonal vectors in ${\mathbb R}^4$
\begin{equation}
\left\{
{\bf w}+{\bf y},
{\bf x}+{\bf z}
\right\} =
\left\{
\frac{1}{2}
\left(\frac{1}{{\sqrt{2}}},-1,
  -\frac{1}{{\sqrt{2}}}  ,{\sqrt{2}}\right),
\frac{1}{2}
\left(  -\frac{1}{2}  ,  -\frac{1}{{\sqrt{2}}}  ,
  -\frac{3}{2} ,-1\right)
\right\}
\label{2003-eu-el2}
\end{equation}
which could be used as a
bit representation.
As can be readily verified, the shares in
(\ref{2003-eu-el1}) are obtained by applying the projections
$P={\rm diag}(1,1,0,0)$
and
$P^\perp={\rm diag}(0,0,1,1)$
to the vectors in (\ref{2003-eu-el2})
[``${\rm diag}(a,b,\ldots )$'' stands for the diagonal matrix with $a,b,\ldots $
at the diagonal entries].
The  comeasurable projection operators corresponding to the vectors
in (\ref{2003-eu-el2}) are given by
\begin{eqnarray}
\left[ ( {\bf w} + {\bf y} )^T ( {\bf w} + {\bf y} ) \right]
&=&
\frac{1}{4}
\left(
\begin{array}{ccccccc}
\frac{1}{2} & - \frac{1}{{\sqrt{2}}}  & - \frac{1}{2}  & 1 \cr
    -\frac{1}{{\sqrt{2}}}   & 1 & \frac{1}{{\sqrt{2}}} & -{\sqrt{2}} \cr
    - \frac{1}{2} & \frac{1}{{\sqrt{2}}} & \frac{1}{2} &-1 \cr
1 & -\sqrt{2} & -1 & 2 \cr
\end{array}
\right)
\qquad {\rm and}\\
\left[ ( {\bf x} + {\bf z} )^T ( {\bf x}+{\bf z} ) \right]
&=&
\frac{1}{4}
\left(
\begin{array}{cccccc}
\frac{1}{4} & \frac{1}{2\, \sqrt{2}} & \frac{3}{4} & \frac{1}{2} \cr
\frac{1}{{2\, \sqrt{2}}} & \frac{1}{2} & \frac{3}{{2\, \sqrt{2}}} & \frac{1}{{\sqrt{2}}}\cr
\frac{3}{4} & \frac{3}{2\, {\sqrt{2}}} & \frac{9}{4} & \frac{3}{2} \cr
    \frac{1}{2} & \frac{1}{{\sqrt{2}}} & \frac{3}{2} & 1 \cr
\end{array}
\right),
\end{eqnarray}
whereas the shares given to the parties are not comeasurable; i.e.,
$
[{\bf w}^T{\bf w}]
[{\bf x}^T{\bf x}]
-
[{\bf x}^T{\bf x}]
[{\bf w}^T{\bf w}]
\neq 0
$, and
$
[{\bf y}^T{\bf y}]
[{\bf z}^T{\bf z}]
-
[{\bf z}^T{\bf z}]
[{\bf y}^T{\bf y}]
\neq 0
$.
Only after recombining the shares it is possible to reconstruct the information;
i.e., to decide whether
$({\bf w}+{\bf y})$
or
$({\bf x}+{\bf z})$
has been communicated.
This configuration demonstrates
the protocol, but it is not optimal, as four dimensions have been used to
represent a single bit.
A more effective coding in base four could utilize the additional two ``quadrit'' states
$(1/2)\left(1,{\sqrt{2}},
  -1 ,0\right)$
and
$(1/2)\left(3/2,-1/{\sqrt{2}},1/2,
  - 1 \right)$.

A possible experimental realization of an arbitrary $m$-dimensional
configuration could be a general interferometer with $m$ inputs and $m$
output terminals \cite{rzbb}, which are partitioned according to the orthogonal projections involved.
They should be arranged in such a way that the single input/output terminals each correspond to one dimension.
Consider, for  example, the two-component two-share configuration discussed above.
The two bit states (\ref{2003-eu-el2}) can be constructed from the orthogonal
pair of vectors
${\bf e}_1 = (0, 0, 0, 1)$ and
${\bf e}_2 = (1, 0, 0, 0)$ by
subjecting them to four successive rotations in two-dimensional subspaces of ${\mathbb R}^4$; i.e.,
${\bf w} + {\bf y} = R_{13}(\pi/4) R_{12}(\pi/4) R_{14}(\pi/4) R_{13}(\pi/4) {\bf e}_1$
and
${\bf x} + {\bf z} = R_{13}(\pi/4) R_{12}(\pi/4) R_{14}(\pi/4) R_{13}(\pi/4) {\bf e}_2$,
where
$R_{12}$,
$R_{14}$,
$R_{13}$
represent the usual clockwise rotations in the
1--2, 1--4, and 1--3 planes.
The corresponding (lossless) interferometric configuration is depicted
in Fig. \ref{2003-eu-f1}; the boxes standing for a 50:50 mixing.
\begin{figure}
\begin{center}
\begin{tabular}{c}
\unitlength 0.85mm
\linethickness{0.8pt}
\begin{picture}(210.00,50.01)
\put(15.00,15.00){\framebox(20.00,30.00)[cc]{$R_{13}({\pi \over 4})$}}
\put(55.00,5.00){\framebox(20.00,40.00)[cc]{$R_{14}({\pi \over 4})$}}
\put(95.00,25.00){\framebox(20.00,20.00)[cc]{$R_{12}({\pi \over 4})$}}
\put(135.00,15.00){\framebox(20.00,30.00)[cc]{$R_{13}({\pi \over 4})$}}
\put(2.00,45.00){\makebox(0,0)[cc]{1}}
\put(2.00,35.00){\makebox(0,0)[cc]{2}}
\put(2.00,25.00){\makebox(0,0)[cc]{3}}
\put(2.00,15.00){\makebox(0,0)[cc]{4}}
\put(0.00,40.00){\vector(1,0){15.00}}
\put(0.00,20.00){\vector(1,0){15.00}}
\put(0.00,10.00){\vector(1,0){55.00}}
\put(0.00,30.00){\vector(1,0){5.00}}
\put(5.00,30.00){\line(2,5){8.00}}
\put(13.00,50.00){\line(1,0){67.00}}
\put(80.00,50.00){\line(1,-4){5.00}}
\put(85.00,30.00){\line(1,0){10.00}}
\put(115.00,30.00){\line(1,0){10.00}}
\put(125.00,30.00){\line(1,3){6.67}}
\put(131.67,50.00){\line(1,0){27.67}}
\put(35.00,40.00){\line(1,0){20.00}}
\put(75.00,40.00){\line(1,0){20.00}}
\put(115.00,40.00){\line(1,0){20.00}}
\put(35.00,20.00){\line(1,0){10.00}}
\put(45.00,20.00){\line(2,-5){8.00}}
\put(53.00,0.00){\line(1,0){26.00}}
\put(79.00,0.00){\line(1,3){6.67}}
\put(85.67,20.00){\line(1,0){49.33}}
\put(159.67,50.00){\line(1,-4){5.00}}
\put(164.67,30.00){\line(1,0){5.33}}
\put(155.00,40.00){\line(1,0){15.00}}
\put(155.00,20.00){\line(1,0){14.67}}
\put(75.00,10.00){\line(1,0){95.00}}
\put(165.00,40.00){\vector(1,0){5.00}}
\put(165.00,30.00){\vector(1,0){5.00}}
\put(165.00,20.00){\vector(1,0){5.00}}
\put(165.00,10.00){\vector(1,0){5.00}}
\bezier{28}(172.33,27.00)(174.67,27.67)(174.67,32.00)
\bezier{20}(174.67,32.00)(175.00,35.00)(177.00,35.00)
\bezier{28}(172.33,43.00)(174.67,42.33)(174.67,38.00)
\bezier{20}(174.67,38.00)(175.00,35.00)(177.00,35.00)
\bezier{28}(172.33,23.00)(174.67,22.33)(174.67,18.00)
\bezier{20}(174.67,18.00)(175.00,15.00)(177.00,15.00)
\bezier{28}(172.33,7.00)(174.67,7.67)(174.67,12.00)
\bezier{20}(174.67,12.00)(175.00,15.00)(177.00,15.00)
\put(183.00,35.00){\makebox(0,0)[lc]{share \# 2}}
\put(183.00,15.00){\makebox(0,0)[lc]{share \# 1}}
\end{picture}
\\
(a)
\\
\\
\unitlength 0.85mm
\linethickness{0.8pt}
\begin{picture}(185.00,50.01)
\put(150.00,15.00){\framebox(20.00,30.00)[cc]{$R_{13}( -{\pi \over 4})$}}
\put(110.00,5.00){\framebox(20.00,40.00)[cc]{$R_{14}(-{\pi \over 4})$}}
\put(70.00,25.00){\framebox(20.00,20.00)[cc]{$R_{12}( -{\pi\over 4})$}}
\put(30.00,15.00){\framebox(20.00,30.00)[cc]{$R_{13}(-{\pi \over 4})$}}
\put(183.00,45.00){\makebox(0,0)[cc]{1}}
\put(183.00,35.00){\makebox(0,0)[cc]{2}}
\put(183.00,25.00){\makebox(0,0)[cc]{3}}
\put(183.00,15.00){\makebox(0,0)[cc]{4}}
\put(170.00,40.00){\vector(1,0){15.00}}
\put(170.00,20.00){\vector(1,0){15.00}}
\put(130.00,10.00){\vector(1,0){55.00}}
\put(180.00,30.00){\vector(1,0){5.00}}
\put(180.00,30.00){\line(-2,5){8.00}}
\put(172.00,50.00){\line(-1,0){67.00}}
\put(105.00,50.00){\line(-1,-4){5.00}}
\put(100.00,30.00){\line(-1,0){10.00}}
\put(70.00,30.00){\line(-1,0){10.00}}
\put(60.00,30.00){\line(-1,3){6.67}}
\put(53.33,50.00){\line(-1,0){27.67}}
\put(150.00,40.00){\line(-1,0){20.00}}
\put(110.00,40.00){\line(-1,0){20.00}}
\put(70.00,40.00){\line(-1,0){20.00}}
\put(150.00,20.00){\line(-1,0){10.00}}
\put(140.00,20.00){\line(-2,-5){8.00}}
\put(132.00,0.00){\line(-1,0){26.00}}
\put(106.00,0.00){\line(-1,3){6.67}}
\put(99.33,20.00){\line(-1,0){49.33}}
\put(25.33,50.00){\line(-1,-4){5.00}}
\put(20.33,30.00){\line(-1,0){5.33}}
\put(30.00,40.00){\line(-1,0){15.00}}
\put(30.00,20.00){\line(-1,0){14.67}}
\put(110.00,10.00){\line(-1,0){95.00}}
\put(20.00,40.00){\vector(1,0){5.00}}
\put(20.00,30.00){\vector(1,0){0.50}}
\put(20.00,20.00){\vector(1,0){5.00}}
\put(20.00,10.00){\vector(1,0){5.00}}
\bezier{28}(12.67,27.00)(10.33,27.67)(10.33,32.00)
\bezier{20}(10.33,32.00)(10.00,35.00)(8.00,35.00)
\bezier{28}(12.67,43.00)(10.33,42.33)(10.33,38.00)
\bezier{20}(10.33,38.00)(10.00,35.00)(8.00,35.00)
\bezier{28}(12.67,23.00)(10.33,22.33)(10.33,18.00)
\bezier{20}(10.33,18.00)(10.00,15.00)(8.00,15.00)
\bezier{28}(12.67,7.00)(10.33,7.67)(10.33,12.00)
\bezier{20}(10.33,12.00)(10.00,15.00)(8.00,15.00)
\put(2.00,35.00){\makebox(0,0)[rc]{share \# 2}}
\put(2.00,15.00){\makebox(0,0)[rc]{share \# 1}}
\end{picture}
\\
(b)
\end{tabular}
\end{center}
\caption{Experimental realization of
(a) the encoding stage of a two-component two-share configuration
by an array of effectively two-dimensional beam splitters depicted as boxes.
The decoding stage (b) is just the encoding stage (a) in reverse order,
with inverse beam splitters. \label{2003-eu-f1}}
\end{figure}
The encoding phase depicted in Fig. \ref{2003-eu-f1}(a) consist of either inserting a particle into the first or the fourth terminal.
Formally,
its state undergoes the particular types of mixing transformations outlined above.
Finally, the two upper and two lower exit terminals
are subdivided into the two shares.
The decoding phase depicted in Fig. \ref{2003-eu-f1}(b) requires both shares, which are
recombined in a reverse interferometric setup,
in which the original states  are reconstructed
by performing the reverse mixings in reverse order.

Some configurations are not usable for secret sharing.
The ``worst case'' scenario might be one in which the first share coincides with
a basis vector of the orthonormal basis spanning ${\mathbb R}^m$. In this case,
the second share just consists of the remaining base states, making possible the detection
of the original message.
Take, for instance, the basis
$
\left\{
(0,0,1),
(0,1,0),
(1,0,0)
\right\}$
which, when projected along the $z$-axis, results in the shares
$
\left\{
(0,0,1)
\right\}$
and
$
\left\{
(0,1,0),
(1,0,0)
\right\}$.
These shares enable the parties to deterministically discriminate
between the first state and the rest (first share), and
between all states (second share).

A simple setup would correspond to a two-dimensional case,
in which a particle
would enter one of two input ports.
A successive beam splitter would then scramble the original signal.
In this setup, the two shares would just correspond to the two output ports of the beam splitter.
Even though both parties would know that the other party would possess a one-dimensional share,
due to phase coherence it would not be possible in a straightforward
manner to reconstruct the secret message
by manufacturing the missing one-component share.

As has already been pointed out, the proposed scheme does not necessarily involve
entangled multipartite states; thus the parties are not given particles as shares.
Rather, in the interferometric realization they are given interferometric channels;
and in order to reconstruct the original message, it is important to
keep quantum coherence among all the parties.
Thus, in the encrypted stage, that is, before the
decoding, no particle detection is allowed,
since this would destroy coherence.
The decoding transformation is the coherent combination of the two
shares whose channels each correspond, respectively, to one and only one secret message.

Here we have proposed to look into possibilities to utilize the higher-dimensional
components of the quantum state by combining
two or more states defined in effectively lower-dimensional subspaces.
Only after all parties have put their parts of the states together, are they
able to decypher the message.
The ``extra dimensions'' not used by the ``flattened out'' subspaces might be very
useful for other purposes as well. For instance,
one might speculate that they  could be exploited for computational
purposes such as speedups; analogously to the introduction of the complex plane
for the solution of certain problems, such as integrals, in analysis.
There, the challenge might be to extend  the existing
quantum algorithms to higher dimensions,
thereby exploiting multidimensional connectedness in search spaces and the like,
and at the same time being able to reconstruct
the results in lower dimensions.


\end{document}